# Determination of Superconducting Gap of SmFeAsF$_x$O$_{1-x}$ Superconductors by Andreev Reflection Spectroscopy

T. Y. Chen[1], S. X. Huang[1], Z. Tesanovic[1], R. H. Liu[2], X. H. Chen[2], and C. L. Chien[1]

[1]*Department of Physics and Astronomy, The Johns Hopkins University, Baltimore, MD 21218 USA*
[2]*Hefei National Laboratory for Physical Science at Microscale and Department of Physics, University of Science and Technology of China, Hefei, Anhui 230026, China*

Abstract

The superconducting gap in FeAs-based superconductor SmFeAs(O$_{1-x}$F$_x$) (x = 0.15 and 0.30) and the temperature dependence of the sample with x = 0.15 have been measured by Andreev reflection spectroscopy. The intrinsic superconducting gap is independent of contacts while many other "gap-like" features vary appreciably for different contacts. The determined gap value of $2\Delta$ = 13.34 ±0.47 meV for SmFeAs(O$_{0.85}$F$_{0.15}$) gives $2\Delta/k_BT_C$ = 3.68, close to the BCS prediction of 3.53. The superconducting gap decreases with temperature and vanishes at $T_C$, in a manner similar to the BCS behavior but dramatically different from that of the nodal pseudogap behavior in cuprate superconductors.

## Introduction

In 1911, after successfully liquefied helium in 1908, Heike Kamerlingh Onnes measured the temperature dependence of the electrical resistivity at low temperatures of mercury (Hg), a metal that could be made very pure by distillation. Instead of confirming one of the several predicted behaviors at that time, nature chose to surprise us all: the resistance of Hg became zero at a finite temperature of $T_C$ = 4.21 K [1]. Ever since 1911 with the discovery of superconductivity in Hg, nature has captivated us with its ever more intriguing superconducting phenomena.

The problem of superconductivity was successfully addressed in 1957 by the Bardeen-Cooper-Schrieffer (BCS) theory [2], which provides an excellent microscopic description of both type I (e.g., Al, Nb, Pb, etc.) and type II (e.g., Nb$_3$Sn, Nb-Ti, etc.) superconductors. These superconductors, the so-called low-$T_C$ conventional superconductors, have s-wave pairing with an isotropic superconducting gap of $2\Delta$ (Fig. 1a), which scales with $T_C$ such that $2\Delta/k_BT_C \approx 3.53$, the well-known BCS result for the s-wave superconductors [2].

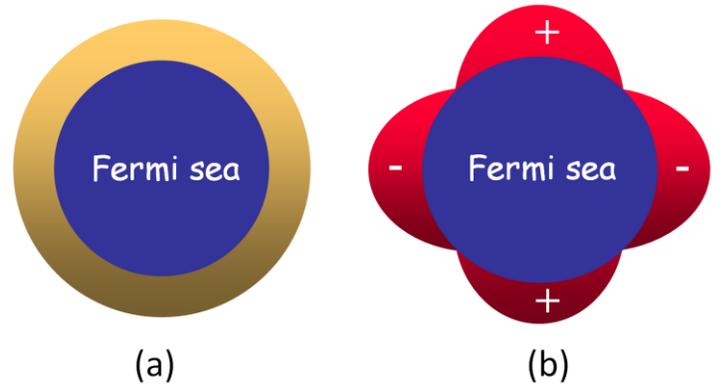

Fig. 1: (a) s-wave pairing in low-$T_C$ conventional and (b) d-wave pairing in high-$T_C$ cuprate superconductors.

The field of superconductivity burst wide open in 1986 with the discovery of cuprate superconductors, first in (La-Sr)$_2$CuO$_4$ ($T_C \approx$ 40 K) [3], followed in quick succession by YBa$_2$Cu$_3$O$_7$ ($T_C \approx$ 90 K) [4], Bi$_2$Sr$_2$Ca$_2$Cu$_3$O$_{10}$ ($T_C \approx$ 105 K), TlBa$_2$Ca$_2$Cu$_3$O$_9$ ($T_C \approx$ 125 K), and HgBa$_2$Ca$_2$Cu$_3$O$_8$ ($T_C \approx$ 135 K). These cuprate superconductors are quasi two-dimensional layered compounds containing the CuO$_2$ planes. The parent compounds (at zero doping) of the cuprates are Mott insulators.

Due to the dominant nearest neighbor antiferromagnetic (AF) interaction, the Cu moments form an AF order with a checkerboard spin structure (Fig. 2a). Upon doping with either electrons or holes, the AF ordering subsides followed by the emergence of the superconducting state, which has unusual superconducting as well as normal state properties. Among the most unusual is the pseudogap, a gap-like feature much larger than the superconducting gap which extends to rather high temperatures [5].

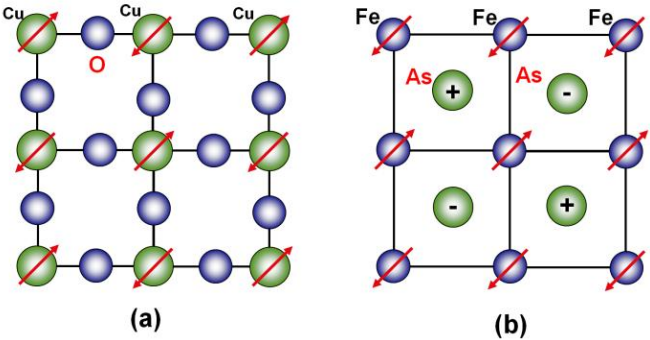

Fig. 2: (a) The $CuO_2$ plane and the AF "checkerboard" ordering in cuprates and (b) the puckered FeAs layer and its SDW ordering in the Fe superconductors.

Unexpectedly, the gaps of the cuprate superconductors were found to be *d*-wave, having the $d_{x2-y2}$ symmetry with the gap vanishing at the four nodes [6] (Fig. 1b), as opposed to *s*-wave in the conventional superconductors. Although a complete understanding is still lacking, it is generally accepted that the intriguing physics of the cuprates is contained in the two-dimensional $CuO_2$ planes, which are essential for the normal state properties, the pseudogap, the *d*-wave pairing, and, of course, high $T_C$. Indeed, it has been a common belief that high-$T_C$ superconductivity is inseparable from *d*-wave pairing.

In March 2008, the field of superconductivity took a dramatic turn with the discovery of the pnictide superconductors (or Fe-based superconductors) [7-12]. These new Fe-pnictide superconductors, with the highest $T_C$ of about 55 K thus far, surprisingly even contain a traditional magnetic element Fe as a crucial ingredient. One immediate response of the scientific community was to cast the new Fe-based superconductors after the familiar cuprate superconductors. There were theoretical predictions [13-18], and experimental confirmations [19-24] of *d*-wave (and also *p*-wave) pairing with nodes in the superconducting gap of these new materials. In contrast, our results using Andreev reflection spectroscopy [25], revealed early on a fully gapped, *s*-wave-like pairing in the FeAs superconductors, a key finding corroborated since in numerous experiments.

The most important quantity of any superconductor is its superconducting gap, which encodes the magnitude, symmetry, and often the intrinsic nature of the pairing mechanism leading to the formation of Cooper pairs. There are relatively few experimental techniques which can be used to either directly measure or indirectly infer the form of the superconducting gap. Andreev reflection [25, 26], angular resolved photon emission spectroscopy (ARPES) [27-31], scanning tunneling microscopy (STM) [5], and tunnel junction [32] are among the techniques that can measure the gap directly, whereas penetration depth [33, 34] and nuclear magnetic resonance (NMR) spin-lattice relaxation rate [35-40] can infer some overall features of the gap, like its being nodal or nodeless.

In this article, we describe the determination of superconducting gap using Andreev reflection spectroscopy, a versatile technique which, among its other attributes, is particularly well-suited for measuring the gap induced by true superconducting correlations rather than other gap structures whose origin is magnetic or structural. Although the Andreev reflection process in the ideal case is well understood, the actual administration of the technique requires quantitative analyses of the differential conductance under various experimental conditions. Most important of all, the success of the technique depends on the ability to differentiate the intrinsic signatures of the superconducting gap from other spurious features caused by the non-ideal contacts.

# Andreev Reflection Spectroscopy

At a normal (N) metal/superconductor (S) interface, a normal current must be converted into a supercurrent, a process in which one electron is accompanied by another electron with opposite spin to form a Cooper pair as required. Consequently, a hole is reflected back into the normal metal, thus doubling the conductance, the famous Andreev reflection process. As a result, the conductance $G(E) = dI/dV$ depends on the energy ($E$) of the injected electrons. Within the gap ($E < \Delta$), $G(E)$ is twice as that outside the gap ($E > \Delta$) as shown by the blue curve in Fig.3 (a). Thus the superconducting gap ($\Delta$) of S can be measured from the energy scan of the conductance of the N/S interface. This is the ideal case for a perfect interface in the clean limit. In reality, an interface is often not ideal. There is interfacial scattering represented by the $Z$ factor, inelastic scattering modeled by the $\Gamma$ factor and the effects of these scattering can be taken into account by the modified Blonder-Tinkham-Klapwijk (BTK) model [41-43]. The interfacial scattering is accommodated by using a $\delta$ potential with strength $Z$ at the interface while the inelastic scattering is incorporated by an imaginary component $i\Gamma$ for the energy. The $Z$ factor suppresses the conductance within the gap as shown in Fig. 3(a) with $Z = 0.00, 0.25, 0.5,$ and $1.0$, while $\Gamma$ affects the conductance mostly near the gap values at $\pm \Delta$ as shown in Fig. 3(b) with $\Gamma/\Delta = 0.0,$ 0.1, 0.2, and 0.5. There is also the effect of finite temperature: the conductance exhibits much sharper features at low temperatures than those at elevated temperatures, such as those shown in Fig. 3 at 4.2 K.

Andreev spectroscopy provides a sensitive and quantitative measurement not only of the gap structure of the superconductor, but also the spin polarization of the metal [42-46] in contact with the superconductor. The conductance within the gap depends sensitively on the spin polarization $P$ of the metal [47]. During the last few years, Andreev spectroscopy has been extensively explored for the measurements of spin polarization of various unpolarized metals, ferromagnetic metals, and half-metals such as $CrO_2$ with 100% spin polarization. The successful application of Andreev spectroscopy has substantiated the technique as a versatile tool for measuring spin polarization and superconducting gap.

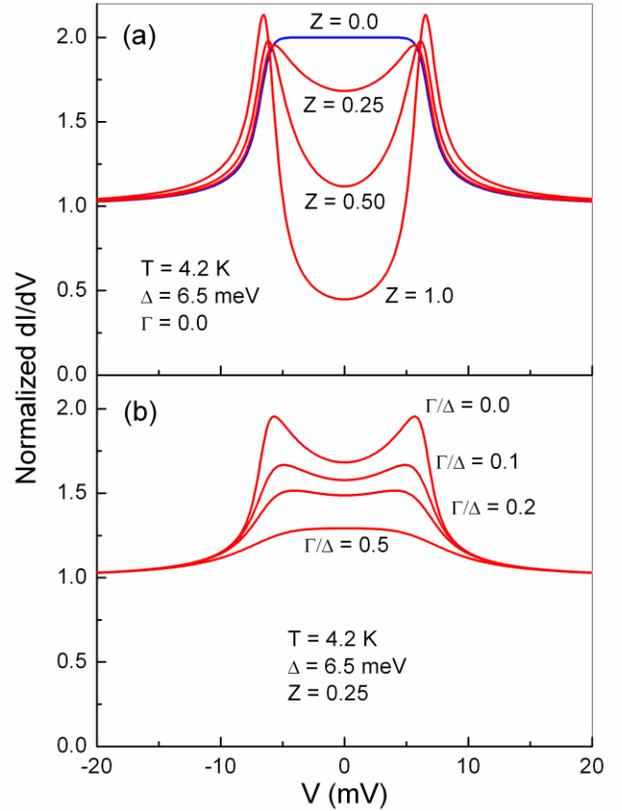

Fig. 3. Andreev spectra computed using the modified BTK model. (a) Spectra for different Z factor: $Z = 0.0$ (blue), 0.25, 0.5, 1.0 with $T = 4.2$ K, $\Delta = 6.5$ meV, $\Gamma = 0.0$, and (b) spectra for different $\Gamma$ factor: $\Gamma/\Delta = 0.0, 0.1, 0.2,$ and $0.5$ with $T = 4.2$ K, $\Delta = 6.5$ meV, $Z = 0.25$.

Andreev spectroscopy is a quantitative technique of measuring conductance intensity as a function of the bias voltage. The bias voltage is theoretically the relative energy between the normal and the superconducting metals. Experimentally, since actual current is being injected at the bias voltage, there are consequences especially under larger bias voltage. Andreev spectroscopy is also quantitative in the intensity of the conductance. For an unpolarized metal such as Cu and Au, the conductance ratio near the gap should be in the vicinity of about 2 (or 100%) as shown in Fig. 3. In the case of a partially polarized ferromagnetic metal, the conductance ratio near the peaks will be reduced accordingly: to about 1.2 for Fe. When the measured conductance is much larger than those allowed by Andreev reflection, e.g., the zero-bias anomaly, it obscures the gap features. By the same token, if the measured conductance is much smaller than that expected from Andreev reflection, one must exercise caution in assigning the meaning of these weaker conductance features. Disagreement among different Andreev reflection measurements often centers on the meaning of the unexpected conductance peaks, whether they are indicative of additional gaps.

The Andreev spectroscopy measurements should be in the ballistic regime without the loss of spin information at the interface. This is accomplished in point contacts with sizes much smaller than the carrier mean free path, where one can use the modified BTK model to analyze the Andreev spectra. When point contact is formed between two materials, the size of the contact can be determined from the measured contact resistance. In the ballistic limit, the contact resistance is the Sharvin resistance $R_C = 4\rho l/3\pi a^2$ [48], where $\rho$ is the resistivity, $l$ the carrier mean free path, and $\pi a^2$ the contact area. Since for Drude metals, $\rho l \approx 10^{-15}$ $m^2$-$\Omega$, one needs contact size $a$ of the order of 50 nm or less to be in the ballistic limit. In Drude metals, the Sharvin formula leads to $[a(nm)]^2 R_C(\Omega) \approx 400$, which sets the scale for the contact resistance in $\Omega$ and contact size in nm. Such small contacts are beyond those achievable by soldering, wire bonding, or even electron beam lithography but can be readily accomplished by mechanical point contacts. The size of such small contacts is usually also much smaller than the crystallite size. Thus, even when a polycrystalline sample is used, the gap value is most often measured from a single grain. The results from multiple contacts on multiple grains should therefore generally hint at whether the gap is quasi isotropic as those in *s*-wave [25] or highly anisotropic such as those in *d*-wave superconductors.

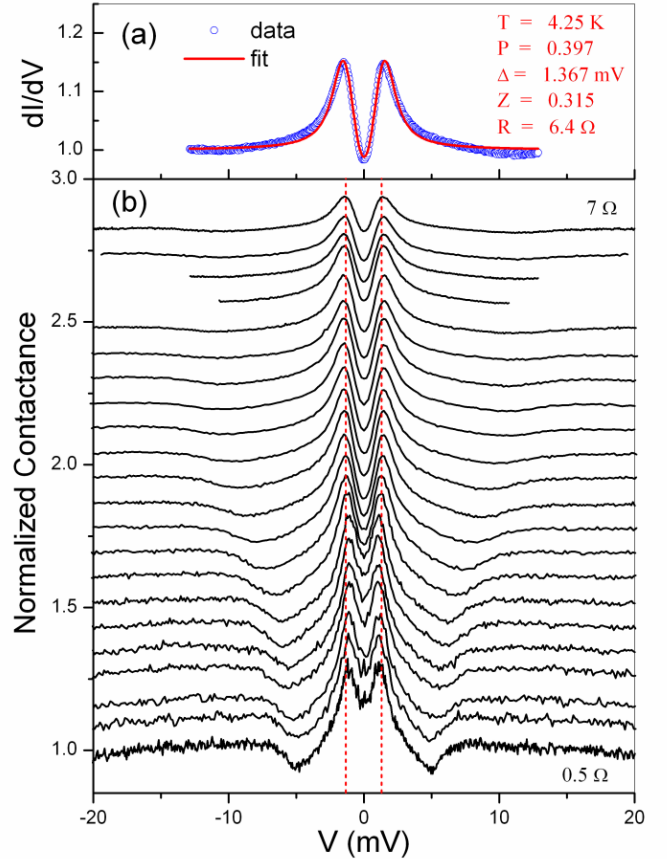

Fig. 4. Andreev spectra of one Nb/Fe contact with contact resistance from 7.0 $\Omega$ to 0.5 $\Omega$. (a) Analysis using the modified BTK model (solid line) of the data (open circles) with contact resistance of 6.4 $\Omega$, and (b) various spectra with contact resistance from 7.0 $\Omega$ to 0.5 $\Omega$. The vertical dashed lines are at ±1.4 mV.

We first illustrate the determination of the superconducting gap using Andreev spectroscopy of a well-known *s*-wave superconductor Nb with an isotropic gap. As illustrated in Fig. 4, using a Nb tip in contact with a bulk Fe sample at 4.2 K with the contact resistance varied from 7.0 $\Omega$ to 0.5 $\Omega$ roughly every 0.3 $\Omega$ or so. When the contact resistance is large, the spectra exhibit the two distinctive Andreev peaks due to the superconducting gap and the spectra can be well described by the modified BTK model as shown in Fig. 4(a), where the open circles are the experimental data and the solid curve is the theoretical fit. The determined gap value for Nb is $\Delta = 1.37$ meV, in excellent agreement with previous reports.

It is important to recognize that the superconducting gap is an intrinsic property of a superconductor and thus must be *independent* of contacts, as clearly demonstrated in Fig. 4(b). As the contact resistance is reduced from 7.0 Ω to 0.5 Ω, the two Andreev peaks remain at the same positions representing the intrinsic gap of $2\Delta$. However, in the conductance spectra a dip also gradually forms outside the gap and it approaches the gap as the contact resistance decreases. If one only had the spectrum with $R_C = 0.5$ Ω, the dip and the "bump" at around 7 mV might be suggestive as a second gap or pseudogap. But a series of conductance spectra shown in Fig. 4(b) illustrate that there is only one superconducting gap. The dip and the associated "bump" are the results of larger contacts and smaller contact resistances. In fact, when $R_C$ is further reduced (data not shown), this dip causes the Andreev peaks to break into multiple peaks and eventually with the appearance of the zero bias anomaly (ZBA). For spectra at small contact resistances (large contact sizes), many other effects such as diffusive transport, critical current density, multiple contacts, non-uniform contacts, and proximity effect can affect the spectra. These complications, which are contact-specific, are not intrinsic features of the superconductors and can be distinguished by measuring the spectrum while altering the contact resistance as shown in Fig. 4(b). Reliable Andreev reflection measurements require not only measurements at multiple contacts but also the ability to slightly vary the contact resistance for each contact. This essential latitude is unfeasible in permanent contacts such as patterned contacts.

**SmFeAsO Superconductors**

The first Fe superconductor that launched the excitement in 2008 has been RFeAsO (where R is one of a few rare earths, e.g., Sm), which embodies many of the salient features of the FeAs superconductors [7-10]. Similar to their cuprate counterparts, the FeAs superconductors are layered compounds. But unlike the cuprates, the $(SmO)^+$ layer and the $(FeAs)^-$ layer in SmFeAsO are substantially puckered, having the Fe atoms form a square lattice with the As atom situated at the centers of the squares, but not in the same plane. From one unit cell to the next, the As atoms are located alternatively above and below the Fe plane (Fig. 2b). The FeAs layer is the prevailing theme appearing in many FeAs superconductors.

The physics of Fe-based superconductors, while apparently distinct from that of conventional and cuprate superconductors, is no less intriguing. Band structure calculations [49, 50] show, and experiments indicate, that there are both electrons and holes at the Fermi surface. The quasi two-dimensional electron and hole pockets form a multiply-connected Fermi surface of which different sections are approximately circular [51]. This situation can be modeled by a tight-binding Hamiltonian which includes all five *d*-orbitals of Fe and their significant overlap with puckered As neighbors [52]. The multiband nature of FeAs layers is prone to the interband pairing mechanism, which can boost superconductivity even if the interband interaction is repulsive [50, 52]. In this case, the superconducting gap will change its sign, with the $+\Delta$ gap for the holes and $-\Delta$ gap for the electrons, while still leaving the Fermi surface fully gapped. In this $s_\pm$ (or *s'*) model, the superconductivity is primarily due to repulsive pairing between the $\pm\Delta$ bands and thus, in principle, does not have to be related to electron-phonon interactions, although such interactions might play a supporting role [52]. The gap features of different Fermi surface pockets have been measured directly by ARPES experiments [27-31], which offer some indication that such sign flipping between the hole and electron pockets is taking place in an otherwise clearly fully gapped system. This unusual quasi two-dimensional band structure thus appears to contain much of the intricate physics of the Fe superconductors.

The parent compound SmFeAsO is a metal and *not* a Mott insulator. This is a crucial difference from the cuprates and it testifies to a significant itinerancy of Fe *d*-electrons [52]. The parent compound is antiferromagnetic with a spin density wave (SDW) ordering below about 140 K (Fig. 2b) due to strong next-neighbor antiferromagnetic interaction but leaving the nearest moments frustrated [11]. The Fe moments, as measured by neutron diffraction, μSR and Mössbauer spectroscopy [53], are anomalously small at about $0.3 - 0.4$ $\mu_B$, only a small fraction of the value expected from the Hund's rule, again in accordance with an itinerant model [52]. Upon doping with F as in $SmFeAsO_{1-x}F_x$, or by means of pressure, the SDW gives way to superconductivity.

In addition to SmFeAsO (the so-called 1111 material), several other FeAs superconductors have been discovered, including $BaFe_2As_2$ (the 122 material) [54-69], in which the $(FeAs)^-$ layers are separated by the $Ba^{2+}$ layers. Superconductivity with $T_C$ up to 38 K can occur by high pressure or by replacing some $Ba^{2+}$ by $K^+$. In LiFeAs, where the $(FeAs)^-$ layers are separated by the $Li^+$ layers [70-74], superconductivity occurs spontaneously, with $T_C \approx 18$ K requiring no doping. It is clear from these examples that because the parent compounds of the Fe superconductors are antiferromagnetic metals, superconductivity can be brought on by doping (e.g., SmFeAsO), by high pressure (e.g., $BaFe_2As_2$), or can occur spontaneously (e.g., LiFeAs). There is also α-FeSe [75], which has a layer structure similar to that of FeAs with $T_C \approx 8$ K.

**Andreev Spectroscopy Measurements of $SmFeAsO_{1-x}F_x$**

We will focus our discussion on $SmFeAsO_{1-x}F_x$ with F doping of x = 0.15 and 0.3. Some samples were fabricated in USTC [8] and others were made at JHU. All the samples are polycrystalline and the values of $T_C$ were obtained from the middle point of the resistivity measurements. We use gold tips in all the measurements. Here we will mainly describe the results for the sample with x = 0.15 and other results will be published elsewhere.

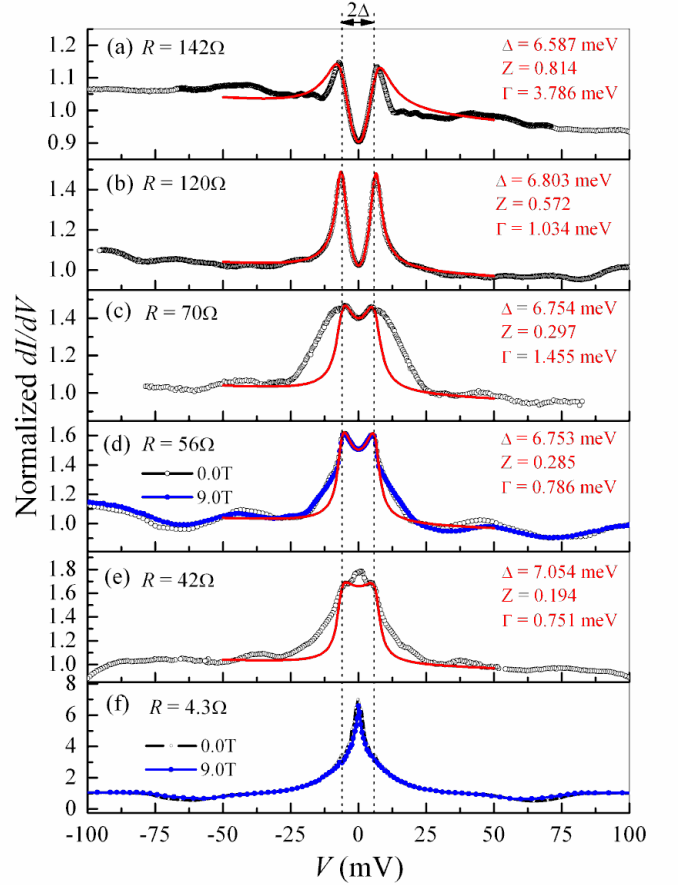

Fig. 5. Andreev spectra of $Au/SmFeAs(O_{0.85}F_{0.15})$ point contacts at 4.52 K with various contact resistance [25]. *a-f*, Spectra are arranged in decreasing contact resistances *R*, where open circles are the experimental data and solid curves are the best fit to the modified BTK model with the parameters and contact resistance listed in each figure. The vertical dashed lines are at ±6.6 mV to indicate the common features. Blue curve in *d* and *f* were results taken in an external magnetic field of *H* = 9 T.

We have measured many Andreev spectra on the sample of $SmFeAs(O_{0.85}F_{0.15})$ with various contact resistances at 4.5 K. The spectra change significantly depending on the contact resistance.

Some representative spectra were plotted in Fig. 5, with contact resistance ranging from 142 Ω to 4.3 Ω. All the spectra show the signature of the Andreev peaks of the gap. Only two peaks were observed (Fig. 5(a-d)) defining a single gap with a value of $\Delta = 6.7$ meV. Using $T_C = 42$ K, we obtain $2\Delta/k_B T_C = 3.68$, close to the value 3.53 for a BCS $s$-wave superconductor.

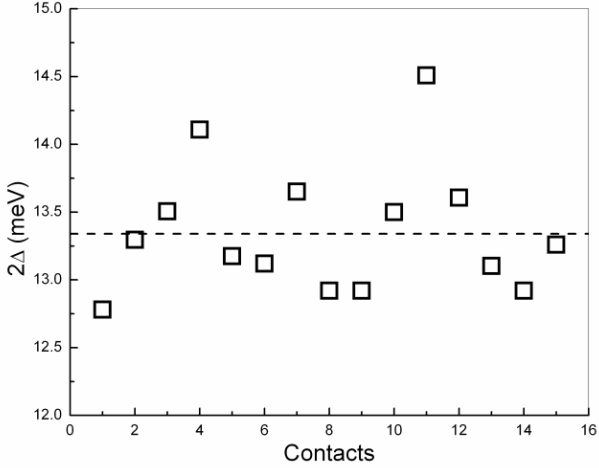

Fig. 6. Gap values $2\Delta$ for different contacts at 4.51 K, and the average value 13.34 meV (dashed line).

In addition to the Andreev peaks due to the superconducting gap, there are small extra features extending to 100 mV as shown in Fig. 5. These extra features, differing from contact to contact, are not intrinsic to the superconductor. Similar to the spectra of Nb shown in Fig. 4, these extra features become more intense at larger contacts with smaller contact resistances. While the origin of these extra features remains complex and elusive, a variety of complications arise for large contact sizes, including diffusive transport, tunneling between grains, magnon and/or phonon excitation, critical current density, and heating. These contact-dependent effects can be identified from the measurements of the Andreev spectra at different contact resistances as shown in Fig. 5.

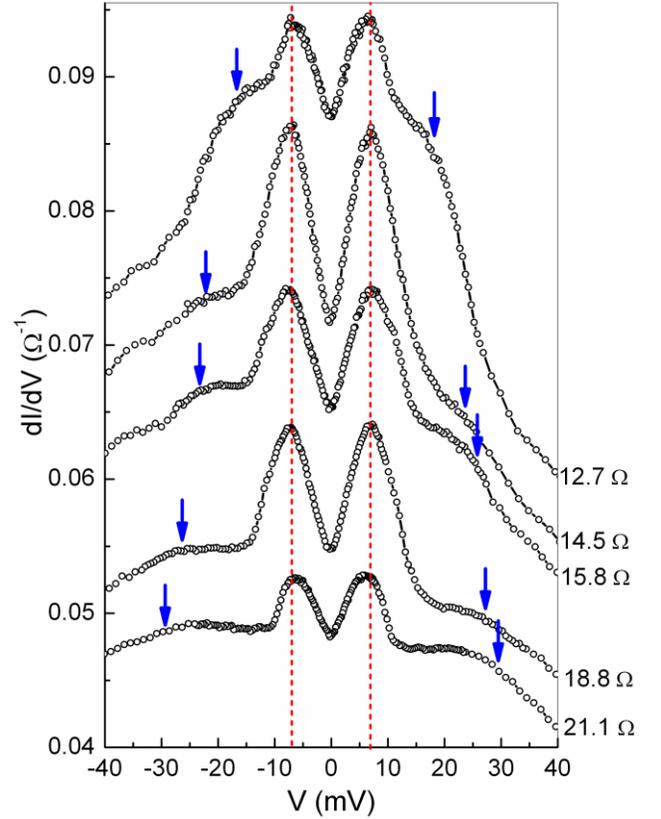

Fig. 7. Andreev spectra of one Au/SmFeAs($O_{0.7}F_{0.3}$) contact with contact resistance gradually tuned from 21.1 Ω to 12.7 Ω. The blue arrows indicate the apparent gap-like features, which change with contact resistance, whereas the actual superconducting gap remain unchanged as shown by the vertical dashed lines.

Because our sample is polycrystalline, each contact most likely connects to only one crystal of a certain orientation. The determined gap values for different contacts are shown in Fig. 6. The average value is $2\Delta = 13.34 \pm 0.37$ meV with a variation less than 10%. Thus the Andreev spectroscopy shows that SmFeAs($O_{0.85}F_{0.15}$) is fully gapped with a distinctively isotropic gap reminiscent of the $s$-wave case, exhibiting at most a 10% anisotropy. Even allowing possible doping variation, the results show that the gap does not exhibit significant anisotropy. Since the Andreev spectra can only measure the magnitude of the gap, we cannot directly address the prospect of two gaps with the similar magnitude but opposite sign as first proposed by Mazin et al. [50] and discussed in [52]. Various features of this unconventional $s_\pm$ or $s'$ superconducting state and methods for its detection have been explored in [76-80].

We further illustrate the importance of varying the contact resistance in the determination of the superconducting gap by Andreev spectroscopy using SmFeAs($O_{0.7}F_{0.3}$) as an example. As shown in Fig. 7, when a contact is established with $R_C$ = 21.1 Ω, it shows the two Andreev peaks at about ±7.3 mV and also two large bumps at about ±30 mV. These latter features in isolation might be taken as evidences for a second gap or a pseudogap. We then slightly decrease contact resistance to 18.8 Ω, and the bump changes to ±27 mV as indicated by the arrow in Fig.7. The bump changes further to ±18 mV at $R_C$ = 12.7 Ω. The fact that these features change with the contact resistance illustrates clearly that they do *not* represent an intrinsic superconducting gap, which must be independent of contact characteristics. Indeed, the two peaks at around ±7.3 mV, which are due to the superconducting gap, as indicated by the red lines, remain unchanged for all the contacts and are evidently intrinsic to the FeAs superconductor. It is interesting to note that the values of $\Delta$ = 7.3 meV and $T_C$ = 53.75 K for SmFeAs($O_{0.7}F_{0.3}$) are higher than $\Delta$ = 6.7 meV and $T_C$ = 42 K for SmFeAs($O_{0.85}F_{0.15}$), but $2\Delta/k_BT_C$ remain close to the value for BCS s-wave superconductors.

Since our report [25] of *s*-wave-like pairing with a $2\Delta/k_BT_C$ value near 3.6, others have come to the same conclusion, although *d*-wave or more exotic pairing symmetry has also been claimed [81 – 83]. We also note instead of a single gap, two gaps have also been claimed from some Andreev reflection measurements on the SmFeAs(O-F) superconductors [83]. Curiously, the lower gap is close to the BCS value whereas the upper gap is considerably larger. Our measurements show, as described in Fig. 7, that only the lower gap is the intrinsic gap.

**Zero Bias Anomalies**

It is sometimes observed, especially with large contact areas, that a spike appears in the conductance *G(V)* at *V = 0*, known as the zero bias anomaly (ZBA). The ZBA can have such a high intensity -- much larger than 2, the limit of ideal Andreev reflection -- that it overwhelms the entire conductance spectrum as shown in Fig. 5(f). The ZBA has been theoretically suggested as a signature of *d*-wave pairing [84]. Indeed, ZBA has been claimed early on as the evidence for *d*-wave pairing in the Fe-pnictide superconductors [20]. However, as we pointed out, the ZBA is not exclusive to *d*-wave; it has been observed in low $T_C$ *s*-wave superconductors such as Al, Pb, $MgB_2$ and Nb as well [85, 86]. Most notably, ZBA appears systematically in the Andreev spectra with smaller contact resistances and thus with larger contact sizes, as shown in Fig. 5, indicating that it is related to the contact geometry and not an intrinsic feature of the superconductor in question. ZBA has also been claimed to be susceptible to magnetic field [20]. Instead, we found the ZBA to be insensitive to magnetic fields up to 9 T at 4.2 K. Regardless of origin of ZBA, once ZBA appears, it overwhelms the Andreev spectra and renders the gap measurements unfeasible.

**Temperature dependence of Andreev Spectrum**

Next, we discuss the temperature dependence of the gap. Fig. 8 shows the Andreev spectra of one point contact on SmFeAs($O_{0.85}F_{0.15}$) from 4.51 K to 56.8 K, every 0.8 K or so. With increasing temperature, both the magnitude and the splitting of the Andreev peaks, hence the gap value, are reduced. The spectrum becomes a single peak and the peak vanishes right at $T_C$ = 42 K. At each temperature, the gap value can be obtained from the theoretical analysis using the modified BTK model. The determined values $2\Delta$ are plotted as solid blue squares in Fig. 9(a).

contact – lack intrinsic physical significance.

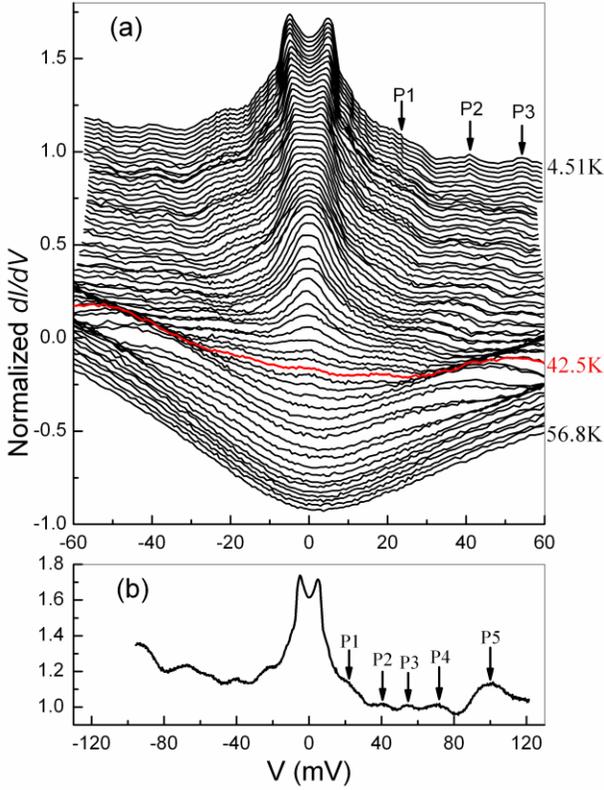

Fig. 8. (a) Andreev spectra from 4.51K to 56.8K of a Au/SmFeAs(O$_{0.85}$F$_{0.15}$) contact with $T_C \approx 42$ K showing the gradual decrease of the gap and peak size, and the eventual disappearance of both at $T_C$, and (b) one Andreev spectrum at 4.51 K measured up to 120 mV, showing minor peaks at P1, P2, P3, P4 and P5.

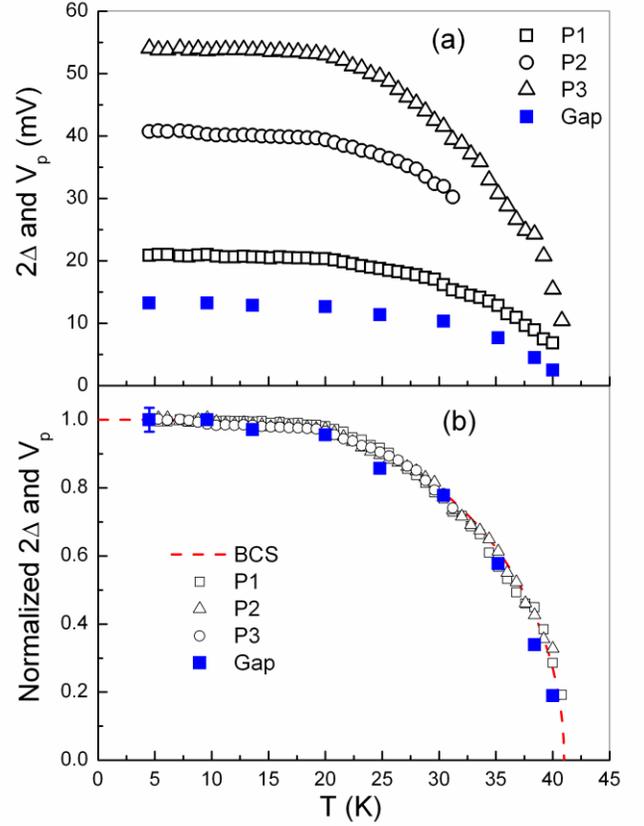

Fig.9. (a) Values of gap 2Δ and peaks indicated in Fig. 8 as P1, P2 and P3, and (b) values of 2Δ, P1, P2, and P3 normalized by those at 4.51 K. The red dashed curve is the BCS s-wave result.

Outside the gap, there are many smaller features over the entire bias voltage range of ±60 mV. With increasing temperature, the extra features outside the gap also reduce to lower bias voltage and eventually vanish at $T_C$. We mark some of the more noticeable extra features by P1, P2, and P3. The temperature dependence of the bias voltage values of the P1, P2, and P3 in Fig. 8 are plotted as open symbols in Fig. 9(a). Both the gap values and peak values of P1, P2, and P3 decrease with increasing temperature and vanish at $T_C$. When normalized, the gap values and peak values of P1, P2, and P3 are all very close to each other and follow the BCS s-wave-like behavior as indicated by the red dashed curve in Fig. 9(b). This further shows that the extra features outside gap are related to the superconductivity, but the features themselves – varying significantly from contact to

The small features outside gap can extend to up to ±120 meV as shown in Fig. 8(b) at 4.51 K. While the actual causes are not well known, these features have been observed in Andreev spectra of conventional *s*-wave superconductors [87, 88] as well as in high quality single-crystal cuprate superconductors [89]. Different mechanisms have been proposed for these features including critical current density [88], bubble effect [87], magnon and/or phonon effects [89], etc. It is worth mentioning here that the point contact has a size of only a few nanometers (which can be estimated from the measured contact resistance using the Sharvin formula) and, with a current of a few mA or a bias voltage of tens of mV, the current density is in the range of $10^8$ to $10^{10}$ A/cm$^2$. This may induce many effects, such as critical current density of the superconductor, Oersted field, and heating, not all of which are intimately tied to the superconductivity. In our data, the peaks P1, P2, and P3 in Fig. 8 appear related to superconductivity since they disappear at right at $T_C$. In contrast, the large peak P5 at about 100 meV persists above $T_C$ and is apparently not related to the superconductivity itself.

Our results using Andreev reflection spectroscopy show that the SmFeAsO superconductors are fully gapped exhibiting a nearly isotropic *s*- or *s'*-wave gap (or potentially some other, more complex fully gapped state like *d+id* or *p+ip*), the result that has been subsequently confirmed by other Andreev reflection measurements [26], ARPES [27-30], and penetration depth studies [33, 34]. Even the NMR spin-lattice relaxation rate results [35-40], the lone dissenter against a nodeless gap in 1111, may be reconciled with the fully gapped picture by invoking the unusual band structure [90].

**Concluding Remarks**

The Andreev reflection spectroscopy allows a quantitative and rather accurate determination of the superconducting gap and its temperature dependence in the newly discovered SmFeAsO materials, as it did in many other superconductors previously. The proper administration of the technique requires multiple contacts, and equally importantly, the ability to alter the contact resistance after a contact has been made. By such exercises, intrinsic *contact-independent* superconducting gap (or gaps) can be determined, with other contact-dependent spurious features identified and discarded.

In the last 98 years since the discovery of superconductivity in Hg, four different types of superconductors have emerged and each has challenged our understanding of the relevant physics. The conventional low-$T_C$ *s*-wave superconductors are governed by electron-phonon interaction, whereas the superconductivity in the heavy-fermion systems is most likely tied to Kondo physics. The *d*-wave superconductivity in the cuprates appears to be governed by the Mott correlations. Most recently, the new Fe-pnictide superconductors are likely to be the examples of the $s_\pm$ (*s'*) interband pairing. Nature continues to dazzle us with intriguing phenomena in superconductivity, with the Fe-pnictides as the latest rendition.

**Acknowledgments**

The authors are grateful to I. Mazin and C. Broholm for useful discussions. This work was supported in part by the NSF grants DMR-0520491 (CLC) and DMR-0531159 (ZT). Work at the Johns Hopkins Institute for Quantum Matter was supported by the U. S. Department of Energy Office of Science under Contract No. DE-FG02-08ER46544 (ZT).

**References:**
[1]. D. van Delft, Physics Today, **61**, 36 (March 2008).
[2]. J. Bardeen, L. N. Cooper, and J. R. Schrieffer, Phys. Rev. **106**, 162; **108**, 1175 (1957).
[3]. J. G. Bednorz and K. A. Müller, Z. Physik B **64**, 189 (1986).
[4]. M. K. Wu, J. R. Ashburn, C. J. Torng, P. H. Hor, R. L. Meng, L. Gao, Z. J. Huang, Y. Q. Wang, and C. W. Chu, Phys. Rev. Lett. **58**, 908 (1987).
[5]. Ø. Fischer, M. Kugler, I. Maggio-Aprile, and C. Berthod, Rev. Mod. Phys. **79**, 353 (2007).

[6]. C. C. Tsuei, J. R. Kirtley, C. C. Chi, Lock See Yu-Jahnes, A. Cgupta, T. Shaw, J. Z. Sun, and M. B. Ketchen, Phys. Rev. Lett. **73**, 593 (1994).
[7]. Y. Kamihara, T. Watanabe, M. Hirano, and H. Hosono, J. Am. Chem. Soc. **130**, 3296 (2008).
[8]. X. H. Chen, T. Wu, G. Wu, R. H. Liu, H. Chen & D. F. Fang, Nature **453**, 761, (2008).
[9]. G. F. Chen, Z. Li, D. Wu, G. Li, W. Z. Hu, J. Dong, P. Zheng, J. L. Luo, and N. L. Wang, Phys. Rev. Lett. **100**, 247002 (2008).
[10]. H.-H. Wen, G. Mu, L. Fang, H. Yang, and X. Zhu, Europhys, Lett. **82**, 17009 (2008).
[11]. C. de la Cruz, Q. Huang, J. W. Lynn, J. Li, W. Ratcliff II, J. L. Zarestky, H. A. Mook, G. F. Chen, J. L. Luo, N. L. Wang and P. Dai, Nature **453**, 899 (2008).
[12]. F. Hunte, J. Jaroszynski, A. Gurevich, D. C. Larbalestier, R. Jin, A. S. Sefat, M. A. McGuire, B. C. Sales, D. K. Christen & D. Mandrus, Nature **453**, 903 (2008).
[13]. K. Kuroki, S. Onari, R. Arita, H. Usui, Y. Tanaka, H. Kontani, and H. Aoki, Phys. Rev. Lett. **101**, 087004(2008)
[14]. P. A. Lee and X.-G. Wen, arXiv: 0804.1739.
[15]. Q. Si and E. Abrahams, Phys. Rev. Lett. **101**, 076401 (2008).
[16]. Z.-J. Yao, J.-X. Li, and Z. D. Wang, arXiv: 0804.4166.
[17]. X.-L. Qi, S. Raghu, C.-X. Liu, D. J. Scalapino and S.-C. Zhang, arXiv: 0804.4332.
[18]. T. A. Maier and D. J. Scalapino, Phys. Rev. **B78**, 020514(R) (2008).
[19]. G. Mu, X. Zhu, L. Fang, L. Shan, C. Ren and H.-H. Wen, Chin. Phys. Lett. 25, 2221-2224 (2008)
[20]. L. Shan, Y. Wang, X. Zhu, G. Mu, L. Fang, and H.-H. Wen, Europhysics Letters 83, 57004 (2008).
[21]. C. Ren, Z.-S. Wang, H. Yang, X. Zhu, L. Fang, G. Mu, L. Shan, and H.-H. Wen, (arXiv: 0804.1726).
[22]. H.-J. Grafe, D. Paar, G. Lang, N. J. Curro, G. Behr, J. Werner, J. Hamann-Borrero, C. Hess, N. Leps, R. Klingeler, B. Büchner, Phys. Rev. Lett. **101**, 047003 (2008)
[23]. P. Samuely, P. Szabó, Z. Pribulová, M. E. Tillman, S. Bud'ko, and P. C. Canfield, *Supercond. Sci. Technol.* **22** No 1 (2009) 014003.
[24]. Y. Wang, L. Shan,_ L. Fang, P. Cheng, C. Ren, and H.-H. Wen, Supercond. Sci. Technol. 22 (2009) 015018
[25]. T. Y. Chen, Z. Tesanovic, R. H. Liu, X. H. Chen, and C. L. Chien, Nature, **453**, 1224 (2008).
[26]. K. A. Yates, L F Cohen, Z.-A. Ren, J. Yang, W. Lu, X.-L. Dong and Z.-X. Zhao, Supercond. Sci. Tech. **21**, 092003 (2008).
[27]. C. Liu, T. Kondo, M. E. Tillman, R. Gordon, G. D. Samolyuk, Y. Lee, C. Martin, J. L. McChesney, S. Bud'ko, M. A. Tanatar, E. Rotenberg, P. C. Canfield, R. Prozorov, B. N. Harmon, and A. Kaminski, arXiv: 0806.2147.
[28]. D. H. Lu, M. Yi, S.-K. Mo, A. S. Erickson, J. Analytis, J.-H. Chu, D. J. Singh, Z. Hussain, T. H. Geballe, I. R. Fisher & Z.-X. Shen, Nature, **455**, 81 (2008)
[29]. W. Malaeb, T. Yoshida, T. Kataoka, A. Fujimori, M. Kubota, K. ono, H. Usui, K. Kuroki, R. Arita, H. Aoki, Y. Kamihara, M. Hirano, and H. Hosono, J. Phys. Soc. Jpn. 77, 093714 (2008).
[30]. T. Kondo, A. F. Santander-Syro, O. Copie, C. Liu, M. E. Tillman, E. D. Mun, J. Schmalian, S. L. Bud'ko, M. A. Tanatar, P. C. Canfield, and A. Kaminski, Phys. Rev. Lett. **101**, 147003 (2008)
[31]. L. Wray, D. Qian, D. Hsieh, Y. Xia, L. Li, J. G. Checkelsky, A. Pasupathy, K. K. Gomes, C. V. Parker, A. V. Fedorov, G. F. Chen, J. L. Luo, A. Yazdani, N. P. Ong, N. L. Wang, and M. Z. Hasan, Phys. Rev. B **78**, 184508 (2008).
[32]. R. Merservey and P. M. Tedrow, Physics Report 238, 173 (1994).
[33]. K. Hashimoto, T. Shibauchi, T. Kato, K. Ikada, R. Okazaki, H. Shishido, M. Ishikado, H. Kito, A. Iyo, H. Eisaki, S. Shamoto, and Y. Matsuda, arXiv: 0806.3149. accepted for publication in Phys. Rev. Lett.
[34]. C. Martin, R. T. Gordon, M. A. Tanatar, M. D. Vannette, M. E. Tillman, E. D. Mun,P. C.

Canfield, V. G. Kogan, G. D. Samolyuk, J. Schmalian, and R. Prozorov, arXiv: 0807.0876.
[35]. Y. Nakai, K. Ishida, Y. Kamihara, M. Hirano, and H. Hosono, Phys. Rev. Lett. **101**, 077006 (2008).
[36]. H.-J. Grafe, D. Paar, G. Lang, N. J. Curro, G. Behr,1 J. Werner, J. Hamann-Borrero, C. Hess, N. Leps, R. Klingeler, and B. Büchner, Phys. Rev. Lett. **101**, 047003 (2008).
[37]. K. Ahilan, F. L. Ning, T. Imai, A. S. Sefat, R. Jin, M. A. McGuire, B. C. Sales, and D. Mandrus, Phys. Rev. **B78**, 100501(R) (2008).
[38]. Y. Nakai1, K. Ishida, Y. Kamihara, M. Hirano,and H. Hosono, J. Phys. Soc. Jpn. **77** (2008) 073701
[39]. K. Matano, Z.A. Ren, X.L. Dong, L.L. Sun, Z.X. Zhao and G.-q. Zheng, Europhys. Lett. 83 (2008) 57001.
[40]. H. Mukuda, N. Terasaki, H. Kinouchi, M. Yashima, Y. Kitaoka, S. Suzuki, S. Miyasaka, S. Tajima, K. Miyazawa, P. Shirage, H. Kito, H. Eisaki, A. Iyo, J. Phys. Soc. Jpn. **77** (2008) 093704
[41]. J. Linder and A. Sudbø, Phys. Rev. **B79**, 020501(R) (2009).
[42]. S. K. Upadhyay, A. Palanisami, R. N. Loui, and R. A. Buhrman, Phys. Rev. Lett. **81**, 3247(1998)
[43]. G. J. Strijkers, Y. Ji, F. Y. Yang, C. L. Chien, and J. M. Byers, Phys. Rev. B **63**, 104510 (2001).
[44]. R. J. Soulen, J. M. Byers, M. S. Osofsky, B. Nadgorny, T. Ambrose, S. F. Cheng, P. R. Broussard, C. T. Tanaka, J. Nowak, J. S. Moodera, A. Barry, and J. M. D. Coey, Science **282**, 85 (1998).
[45]. Y. Ji, G. J. Strijkers, F. Y. Yang, C. L. Chien, J. M. Byers, A. Anguelouch, G. Xiao, and A. Gupta, Phys. Rev. Lett. **86**, 5585 (2001).
[46]. I. I. Mazin, A. A. Golubov, and B. Nadgorny, J. Appl. Phys. **89**, 7576 (2001).
[47]. J. M. D. Coey and C. L. Chien, MRS Bulletin **28** (no.10), 720 (October 2003).
[48]. G. Wexler, Proc. Phys. Soc. **89**, 927-941 (1966).
[49]. D. J. Singh and M. H. Du, Phys. Rev. Lett. **100**, 237003 (2008).
[50]. I. I. Mazin, D. J. Singh, M. D. Johannes, and M. H. Du, Phys. Rev. Lett. **101**, 057003 (2008).
[51]. A.I. Coldea, J.D. Fletcher, A. Carrington, J.G. Analytis, A.F. Bangura, J.-H. Chu, A. S. Erickson, I.R. Fisher , N.E. Hussey, and R.D. McDonald, Phys. Rev. Lett. **101**, 216402 (2008).
[52]. V. Cvetkovic and Z. Tesanovic, arXiv: 0804.4678.
[53]. H. Luetkens, H.-H. Klauss, R. Khasanov, A. Amato, R. Klingeler, I. Hellmann, N. Leps, A. Kondrat, C. Hess,A. Köhler, G. Behr, J. Werner, and B. Büchner, Phys. Rev. Lett. **101**, 097009 (2008).
[54]. M. Rotter, M. Tegel, and D. Johrendt, Phys. Rev. Lett. **101**, 107006 (2008).
[55]. G. Li, W. Z. Hu, J. Dong, Z. Li, P. Zheng, G. F. Chen, J. L. Luo, and N. L. Wang, Phys. Rev. Lett. **101**, 107004 (2008).
[56]. K. Sasmal, B. Lv, B. Lorenz, A. M. Guloy, F. Chen, Y.-Y. Xue, and C.-W. Chu, Phys. Rev. Lett. **101**, 107007 (2008).
[57]. A. S. Sefat, R. Jin, M. A. McGuire, B. C. Sales, D. J. Singh, and D. Mandrus, Phys. Rev. Lett. **101**, 117004 (2008).
[58]. N. Ni, S. L. Bud'ko, A. Kreyssig, S. Nandi, G. E. Rustan, A. I. Goldman, S. Gupta, J. D. Corbett, A. Kracher, and P. C. Canfield, Phys. Rev. **B78**, 014507 (2008).
[59]. M. Rotter, M. Tegel, and D. Johrendt, Phys. Rev. **B78**, 020503(R) (2008).
[60]. J.-Q. Yan, A. Kreyssig, S. Nandi, N. Ni, S. L. Bud'ko, A. Kracher, R. J. McQueeney, R. W. McCallum, T. A. Lograsso, A. I. Goldman, and P. C. Canfield, Phys. Rev. **B78**, 024516 (2008).
[61]. Z. Ren, Z. Zhu, S. Jiang, X. Xu, Q. Tao, C. Wang, C. Feng, G. Cao, and Z. Xu, Phys. Rev. **B78**, 052501 (2008).
[62]. H. S. Jeevan, Z. Hossain, Deepa Kasinathan, H. Rosner, C. Geibel, and P. Gegenwart, Phys. Rev. **B78**, 052502 (2008).

[63]. A. P. Litvinchuk, V. G. Hadjiev, M. N. Iliev, Bing Lv, A. M. Guloy, and C. W. Chu, Phys. Rev. **B78**, 060503(R) (2008).

[64]. C. Krellner, N. Caroca-Canales, A. Jesche, H. Rosner, A. Ormeci, and C. Geibel, Phys. Rev. **B78**, 100504(R) (2008).

[65]. A. I. Goldman, D. N. Argyriou, B. Ouladdiaf, T. Chatterji, A. Kreyssig, S. Nandi, N. Ni, S. L. Bud'ko, P. C. Canfield, and R. J. McQueeney, Phys. Rev. **B78**, 100506(R) (2008).

[66]. R. Mittal, Y. Su, S. Rols, T. Chatterji, S. L. Chaplot, H. Schober, M. Rotter, D. Johrendt, and Th. Brueckel, Phys. Rev. **B78**, 104514 (2008).

[67]. G. Mu, H. Luo, Z. Wang, L. Shan, C. Ren and H.-H. Wen, (arXiv: 0808.2941).

[68]. H. Ding, P. Richard, K. Nakayama, T. Sugawara, T. Arakane, Y. Sekiba, A. Takayama, S. Souma, T. Sato, T. Takahashi, Z. Wang, X. Dai, Z. Fang, G. F. Chen, J. L. Luo, and N. L. Wang, Europhysics Letters 83, 47001 (2008)

[69]. C. Liu, G. D. Samolyuk, Y. Lee, N. Ni, T. Kondo, A. F. Santander-Syro, S. L. Bud'ko, J. L. McChesney, E. Rotenberg, T. Valla, A. V. Fedorov, P. C. Canfield, B. N. Harmon, and A. Kaminski, Phys. Rev. Lett. 101, 177005 (2008)

[70]. J. H. Tapp, Z. Tang, B. Lv, K. Sasmal, B. Lorenz, Paul C. W. Chu, and A. M. Guloy, Phys. Rev. **B78**, 060505(R) (2008).

[71]. D. J. Singh, Phys. Rev. **B78**, 094511 (2008).

[72]. X. C. Wang, Q. Q. Liu, Y. X. Lv, W. B. Gao, L. X. Yang, R. C. Yu, F. Y. Li, C. Q. Jin, arXiv: 0806.4688.

[73]. I. A. Nekrasov, Z. V. Pchelkina, and M. V. Sadovskii, JETP Letters, 2008, Vol. 88, No. 8, pp. 543-545

[74]. M. J. Pitcher, D. R. Parker, P. Adamson, S. J. C. Herkelrath, A. T. Boothroyd and S. J. Clarke, Chemical Communications 5918 - 5920 (2008)

[75]. F.-C. Hsu, J.-Y. Luo, K.-W. Yeh, T.-K. Chen, T.-W. Huang, P. M. Wu, Y.-C. Lee, Y.-L. Huang, Y.-Y. Chu, D.-C. Yan, and M.-K. Wu, arXiv: 0807.2369.

[76]. A. V. Chubukov, D. Efremov, and I. Eremin, Phys. Rev. **B 78,** 134512 (2008).

[77]. V. Cvetkovic and Z. Tesanovic, arXiv:0808.3742.

[78]. M. Parish, J. Hu, and B. A. Bernevig, Phys. Rev. **B 78**, 144514 (2008).

[79]. V. Stanev, J, Kang, and Z. Tesanovic, Phys. Rev. **B 78**, 184509 (2008).

[80]. P. Ghaemi, F. Wang, and A. Vishwanath, arXiv:0812.0015.

[81]. K. A. Yates, L. F. Cohen, Z. A. Ren, J. Yang, W. Lu, X. L. Dong, and Z. X. Zhao, Supercond. Sci. Technol. 21, 092003 (2008).

[82]. O. Millo, T. Asulin, O. Yuli, I. Felner, Z. A. Ren. X. L. Shen, G. C. Che, Z. X. Zhao, Phys. Rev. B 78, 092505 (2008).

[83]. D. Daghero, M. Tortello. R. S. Gonnelli, V. A. Stepanov, N. D. Zhigadlo, and Karpinski, arXiv: 0812.1141.

[84]. G. Deutscher, Rev. Mod. Phys., **77**, 109-135 (2005).

[85]. G. Li, H. He, Y. Wang, L. Lu, S. Li, X. Jing, D. Zhang, Phys. Rev. Lett. **82**, 1229-1232 (1999).

[86]. P. Xiong, G. Xiao, and R. B. Laibowitz, Phys. Rev. Lett. 71, 1907-1910 (1993).

[87]. A. Hahn and K. Hümpfner, Phys. Rev. B 51, 3660 (1995).

[88]. P. S. Westbrook and A. Javan, Phys. Rev. **B 59**, 14606(1999).

[89]. N. Achsaf, G. Deutscher, A. Revcolevschi, and M. Okuya, 1996, Coherence in High Temperature Superconductors, edited by G. Deutscher and A. Revcolevschi (World Scientific, Singapore), p. 428.

[90]. D. Parker, O.V. Dolgov, M.M. Korshunov, A.A. Golubov, and I.I. Mazin, Phys. Rev. B 78, 134524 (2008)